\documentclass[preprint,12pt]{elsarticle}
\usepackage{amssymb}

\journal{Optics Communications}

\begin{document}

\begin{frontmatter}

\title{Trapping of Atoms by the Counter-Propagating Stochastic Light
Waves}

\author{V. I. Romanenko}
\cortext[cor1]{V. I. Romanenko}
\ead{vr@iop.kiev.ua}
\author{L. P. Yatsenko}
\address{Institute of Physics of the National Academy of
  Sciences of Ukraine, 46, Nauky Ave., Kyiv 03680, Ukraine}

\begin{abstract}

  We {calculate} the temperature of the atoms in the
  field of counter-pro\-pa\-ga\-ting stochastic light waves (the
  chaotic-field model). {We show that the} temperature of the atomic
  ensemble depends on the autocorrelation time of the waves, their
  intensity and the detuning of the carrier frequency of the waves
  from the atomic transition frequency. The field can form a
  {one-dimensional} trap for atoms, {as} is readily seen from our
  previous investigation of light-pressure force on an atom in
  counter-propagating stochastic light waves [V.~I.~Romanenko,
  B. W. Shore, L. P. Yatsenko, Opt.\ Commun. 268 (2006) 121--132].
  {We carry out} numerical simulation of the atomic ensemble {using
    paramaters appropriate} for sodium atoms. Analyzing the known
  investigation of the light-pressure force on atoms and their motion
  in the counter-propagating polychromatic waves, we suggest {an}
  hypothesis that any polychromatic counter-propagating waves {that
    have a} discrete spectrum, or waves described by a stationary
  stochastic process, one of which repeats the other, can form a trap
  for atoms.

\end{abstract}


\end{frontmatter}

\section{Introduction}
\label{sec:introduction}

Optical traps for atoms in which, in addition to the confinement, the
atoms are cooled, are widely used in experiments with cold atoms.  For
example, confinement and simultaneous cooling of atoms is realized in
a widely used magneto-optical trap~\cite{Metcalf}, {wherein} the atoms
are subjected to laser radiation and a magnetic field.  Recently,
after a series of articles discussing the formation of a trap for
atoms by the trains of counter-propagating light pulses, one of which
repeats the
other~\cite{Freegarde,Goepfert,Romanenko-JPB,Balykin,Yanyshev}, it was
shown that such a trap can also cool atoms, provided that the pulse
parameters are properly
chosen~\cite{Romanenko-UJP-2013,Romanenko-JMO,Romanenko-PRA}.  Another
trap that simultaneously confines and cools atoms, but {which is}
based on two collinear standing waves {that} can be treated as
counter-propagating bichromatic light waves, was proposed
in~\cite{Voitsekhovich} and investigated by the authors in the recent
paper~\cite{Romanenko-UJP-2016}.  The {centers} of both types of trap,
the trap based on the counter-propagating light pulses and the trap
based on the counter-propagating bichromatic waves, are situated at
the point where the optical paths of the counter-propagating waves,
produced by the same laser, are equal.

\begin{figure}
\centerline{\includegraphics[]{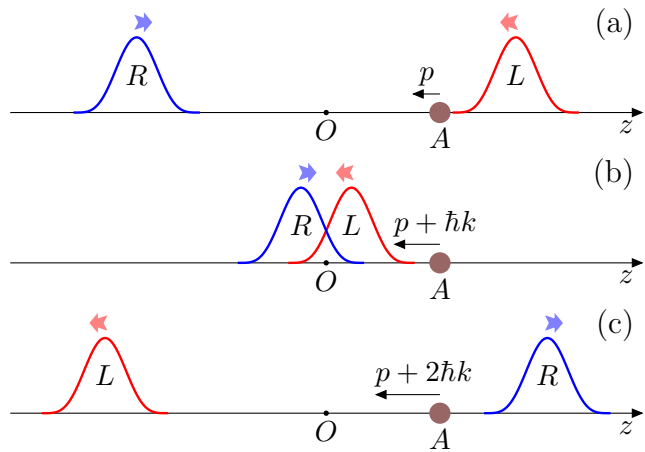}}
\caption{Interaction of a two-level atom (A) with counter-propagating
  $\pi$-pulses.  {(a)} Initially the atom occupies the ground state
  and its momentum is $p$.  {(b)} The pulse $L$, propagating {in the
    negative} $z$ direction, excites the atom and its momentum becomes
  $p+\hbar k$, where $\hbar k$ is the momentum of the photon.  {(c)}
  Interaction with the counter-propagating pulse $R$ leads to the
  stimulated emission of a photon in the direction of $z$ axis and
  additional change of the atomic momentum by $\hbar k$. As a result,
  the atomic momentum is changed by $2\hbar k$ after interaction with
  each pair of counter-propagating pulses. {The momentum change is
    directed toward} the point $O$ where the pulses ``collide''}
\label{fig-1}
\end{figure}

The idea of {a} trap formed by the counter-propagating trains of light
pulses can be simply explained for the case of
$\pi$-pulses~\cite{Freegarde}. Let's consider a two-level atom $A$ {at a}
 point where the pulses {do} not overlap (Figure~\ref{fig-1}).
The atom interacts with a $\pi$-pulse $L$ {traveling leftward toward}
the point $O$ where the pulses ``collide''.  {After a} time delay
shorter then the spontaneous emission time $\tau_{sp}$ {it} interacts
with the $\pi$-pulse $R$ propagating in the opposite direction. The
first pulse excites the atom and pushes it {toward} the point $O$ due
to the absorption of a photon.  The interaction with the second pulse
leads to stimulated emission of the photon in the direction of its
propagation, pushing the atom {again toward} point $O$.  If the pulse
repetition period is large ($T\gg\tau_{sp}$), the atom always occupies
the ground state before the interaction with a counter-propagating
pair of pulses and a trapping force is provided. More detailed
consideration shows that the trap can be formed even by pulses with
{an} area much smaller {than} $\pi$ and the condition $T\gg\tau_{sp}$
is not obligatory~\cite{Romanenko-PRA}.  It should be {noted} that the
light-pressure force on the atoms in the field of counterpropagating
bichromatic waves (``bichromatic force''~\cite{soding1997short}) can
be qualitatively interpreted as {an} interaction with
counter-propagating light pulses~\cite{Voitsekhovich,soding1997short}.
Both the bichromatic force and the force in the field of the
counter-propagating laser pulses {is} equal to zero {at} the point
where the optical paths of the counter-propagating waves, produced by
the same laser, are equal (the center of the trap).  Near this point
the light-pressure force depends {almost linearly} on the coordinate,
{thereby} providing the restoring force for any deviation of an atom
from the center of the trap.  The force arises because the field
{strengths} of the counter-propagating waves are correlated (in the
opposite case the averaged force would be zero, as was shown
in~\cite{Romanenko-OC} for stochastic counter-propagating waves with
time delay between waves more than the autocorrelation
time). Therefore, we can expect the existence of the restoring force
even in the case of the stochastic field, provided the autocorrelation
time of the waves is much longer {than} the time of light propagation
from the position of the atom to the center of the trap. This
consideration is in good agreement with the calculation of the
light-pressure force acting on an atom in counter-propagating
stochastic waves~\cite{Romanenko-OC}.

The existence of the restoring force is a necessary but not sufficient
condition for confinement of atoms in {a} trap.  The laser radiation
heats the atoms due to momentum diffusion from the scattering of laser
photons. In optical traps for atoms the heating is compensated by a
``friction force'' which originates from the interaction of atoms with
monochromatic standing waves {that are} slightly detuned from the
atomic transition~\cite{Metcalf}. In the case of the
counter-propagating stochastic wave we should {determine whether} the
continuous spectrum of radiation with spectral width more then
$1/\tau_{sp}$ can cool atoms and how the spectral width affects the
temperature of the possible cooling.  In this paper we examine the
interaction of atoms with counter-propagating stochastic plane waves
(a chaotic-field model, where the real and imaginary parts of the
complex amplitude of the electrical field {fluctuate} independently),
and show that in this case, as in the case of the counter-propagating
laser pulses or bichromatic field, the atoms can be confined and
cooled by the same field.  The trap under consideration combines the
idea of the confinement of atoms by the counter-propagating waves and
their cooling by white light~\cite{Hoffnagle,Park,Glover} using the
same laser radiation.

This paper is organized as follows.  In Section~\ref{sec:trap-model}
we present the trap model {considered} in this paper. In Section
\ref{sec:main-equations} the equations that describe the evolution of
the atom in the stochastic field are written, in
Section~\ref{sec:atoms-weak-field} the light-pressure force on an atom
in the weak field is derived and the minimal temperature of the atomic
ensemble in the stochastic field is found.  In
Section~\ref{sec:numer-calc-proc} we describe the numerical
calculation routine used in the investigation. Results of the
numerical calculations of the statistic properties of the ensemble of
sodium atoms and their discussion are presented in
Section~\ref{sec:results-discussion}.  The conclusions are given in
Section~\ref{sec:conclusions}.  In the appendix, we explain the origin
of the momentum diffusion of atoms in the field of laser radiation.

\section{Trap model}
\label{sec:trap-model}

According to the results of~\cite{Romanenko-OC}, the light pressure
force in the field of the counter-propagating stochastic waves, one of
which repeats the other with a certain time delay, is directed to the
point where this delay is equal to zero. Thus, it is possible to form
an optical trap by the stochastic light field. {Whenever} the
radiation field of a multimode laser is close to
stochastic~\cite{Georges79,Georges80}, it is possible to construct a
trap on the basis of such lasers.

\begin{figure}
\centerline{\includegraphics[]{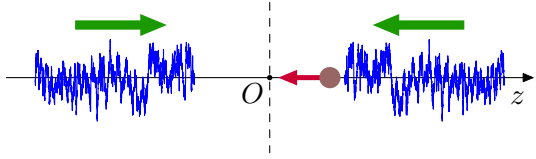}}
\caption{{Schematic view} of the trap. The atom (indicated by the
  circle) near the point $O$ is subjected to the field of the
  counter-propagating light waves with the stochastic {envelope}. The
  waves are produced by the same source and in the point $O$ their
  amplitudes and phases are equal. Due to the force of light pressure
  the atom moves near the center $O$ of the trap}
\label{fig-2}
\end{figure}

A {schematic drawing} of an one-dimensional trap for atoms, formed by the
counter-propagating stochastic waves is depicted in
Figure~\ref{fig-2}.
The waves are produced by the same laser and {are} directed towards each
other by a system of mirrors (not shown in the figure).

\section{Main equations}
\label{sec:main-equations}

In the calculations we assume the condition~\cite{Minogin,Dalibard94}
\begin{equation}
  \label{eq:crit}
\frac{\hbar^{2}k^{2}}{2m}\ll\hbar\gamma,  
\end{equation}
which means that the light-pressure force is formed faster than the
change of the atomic velocity will have a significant impact on its
value (the heavy atom approximation).  Here $k=\omega/c$ is the wave
vector, $\omega$ is the carrier frequency of the waves, $\gamma$ is
the rate constant of spontaneous emission. Criterion~(\ref{eq:crit})
is the basis for the semi-classical approach in derivation of the
Doppler-cooling limit~\cite{Dalibard94}.

Let's consider an atom in the field of two counter-propagating waves,
one of which repeats the other with some time delay. The origin of the
coordinate system is situated {at} the point $O$ (see
figure~\ref{fig-2}), where this delay is equal to zero.  The atom with
the coordinate $z$ is subjected to the electric field
\begin{equation}
 \mathbf{E}
=\frac{1}{2}\mathbf{e}
\bigl[E_{0}(t-z/c)\exp\left(i\omega t-i kz\right)
+E_{0}(t+z/c)\exp\left(i\omega t+i kz\right)\bigr] +\mbox{ c.\,c.}
  \label{eq:E}
\end{equation}
Here $\mathbf{e}$ is the unit vector of polarization.

We describe the laser radiation by the chaotic field
model~\cite{Georges79,Georges80}, in which the real and imaginary
parts of the complex amplitude of the field fluctuate
independently. The ensemble average of the complex amplitude equals to
zero and the autocorrelation functions of the real and the imaginary
parts are
\begin{eqnarray}
  \langle\mathop{\mathrm{Re}}E_{0}(t)\mathop{\mathrm{Re}}E_{0}(t')\rangle
  &=&\frac{1}{2}|E_{0}|^{2} \exp{-G|t-t'|}),
\label{eq:E-chaotic_re}
\\
  \langle\mathop{\mathrm{Im}}E_{0}(t)\mathop{\mathrm{Im}}E_{0}(t')\rangle
  &=&\frac{1}{2}|E_{0}|^{2} \exp({-G|t-t'|}),
\label{eq:E-chaotic_im}
\end{eqnarray}
where brackets $\langle\quad\rangle$ denote averaging over the
ensemble of the possible realizations of the stochastic process, $G$
is the inverse correlation time and
$|E_{0}|^{2}=\langle E_{0}(t) E_{0}^{*}(t)\rangle$ does not depend on
time.

We use a simple two-level model of the atom-field interaction and
assume the absence of polarization gradients.  The difference of
energies of the ground $\left| 1\right\rangle$ and the excited
$\left| 2\right\rangle$ states is $\hbar\omega_{0}$. The light
pressure force on the atom is given by the
formula~\cite{Metcalf,Minogin}
\begin{equation}
  F=(\varrho_{12}\mathbf{d}_{21}+
  \varrho_{21}\mathbf{d}_{12})\frac{\partial \mathbf{E}}{\partial z},
\label{eq:F}
\end{equation}
where $\mathbf{d}_{12}$ and $\mathbf{d}_{21}$ are the matrix elements
of the dipole moment, $\varrho_{12}$ and $\varrho_{21}$ are the
non-diagonal elements of the density matrix $\varrho$.

According to~(\ref{eq:F}) {a force calculation requires we know not
  only the fields but also} the density matrix of the atom.  We use
two approaches for calculating the density matrix of the atom.  In our
analytic calculation of the temperature of the atomic ensemble in a
weak field we use the density matrix equation. In our numerical
calculation of the statistical properties of the atomic ensemble we
use the Monte Carlo wave-function method~\cite{Mol93}. The results of
both methods are compared in Section~\ref{sec:results-discussion} and
shows good agreement.

Under the action of force~(\ref{eq:F}) the atom {accelerates} according to
Newton's law
\begin{equation}
\dot{v}=F/m,\label{eq:ma}
\end{equation}
where $m$ is the mass of the atom and $v=\dot{z}$ is its velocity.

\section{Atoms in a weak field}
\label{sec:atoms-weak-field}
To explain laser cooling by {a} stochastic field, we will find the
light pressure force (\ref{eq:F}) on an atom and the population of the
excited state in a weak field ($\Omega_{0}\ll\gamma$) using the
equation for the density matrix
\begin{equation}
  \label{eq:dm}
  i\hbar\frac{\partial}{\partial t}\varrho_{jk}=\sum_{l}(H_{jl}\varrho_{lk}- 
  \varrho_{jl}H_{lk})+i\hbar\sum_{l,m}\Gamma_{jk,lm}\varrho_{lm},  
\end{equation}
where the Hamiltonian $H$ is equal to
\begin{equation} 
  \label{eq:Hamdm}
  {{H}}= \hbar\omega_{0}|2\rangle\langle2|-
  \mathbf{d}_{12}|1\rangle\langle2|\mathbf{E}(t)-
  \mathbf{d}_{21}|2\rangle\langle1|\mathbf{E}(t).
\end{equation}
The relaxation process in (\ref{eq:dm}) is described by the term
containing the {array of values} $\Gamma$. {The non-zero} elements of
$\Gamma$ are
\begin{eqnarray}
  \label{eq:Gamma}
\Gamma_{12,12}&=&\Gamma_{21,21}=-\gamma/2,\nonumber\\    
\Gamma_{11,22}&=&-\Gamma_{22,22}=\gamma.
\end{eqnarray}
After substitution of (\ref{eq:Hamdm}), (\ref{eq:Gamma})
and~(\ref{eq:E}) in (\ref{eq:dm}), applying rotating wave
approximation~\cite{Shore} and {introducing} the variables
\begin{equation}
  \label{eq:sigma}
\sigma_{12}={}  \varrho_{12}e^{-i\omega_{0}t},\qquad  
\sigma_{21}  =\varrho_{21}e^{i\omega_{0}t}
\end{equation}
we get
\begin{eqnarray}
 \frac{\partial}{\partial
      t}\varrho_{11}&=&\frac{i}{2}\sigma_{12}e^{i\delta t}
    \left[
      \Omega_{1}^{*}e^{i kz} +\Omega_{2}^{*}e^{-i kz}
    \right]-\frac{i}{2}\sigma_{21}e^{-i\delta t}
    \left[
      \Omega_{1}e^{-i kz} +\Omega_{2}e^{ ikz}
    \right]\nonumber\\
    &&+\gamma\varrho_{22},    \label{eq:dmrwa-i}\\
    \frac{\partial}{\partial
      t}\sigma_{12}&=&\frac{i}{2}(\varrho_{11}-\varrho_{22})e^{-i\delta t}
    \left[
      \Omega_{1}e^{-i kz}\right. 
   \left.+\Omega_{2}e^{i kz}
    \right]-\frac{\gamma}{2}\sigma_{12}, \label{eq:dmrwa-ii}\\
    \sigma_{21}&=&\sigma_{12}^{*},\qquad \varrho_{11}+\varrho_{22}=1,
  \label{eq:dmrwa-iii}
\end{eqnarray}
where $\Omega_{1}=\Omega(t-z/c)$, $\Omega_{2}=\Omega(t+z/c)$,
$\Omega(t)=-\mathbf{d}_{12}\mathbf{e}E_{0}(t)/\hbar$,
$\delta=\omega_{0}-\omega$.  The light-pressure force in this
approximation is equal to

\begin{equation}
  F=\hbar k \frac{i}{2}e^{i\delta t}\sigma_{12}
  \left[
    \Omega_{2}^{*}e^{-i kz}-\Omega_{1}^{*}e^{i kz}
  \right]-\hbar k \frac{i}{2}e^{-i\delta t}\sigma_{21}
  \left[
    \Omega_{2}e^{i kz}-\Omega_{1}e^{-i kz}
  \right].
  \label{eq:frwa}
\end{equation}

We {seek} the quasi-stationary solution to the equations
(\ref{eq:dmrwa-i})--(\ref{eq:dmrwa-iii}). {This} is achieved when
$t\gg\gamma^{-1}$. Introducing the parameter $\varepsilon$ to denote
the order of smallness of $\Omega_{1}$, $\Omega_{2}$ in comparison
with $\gamma$ (after the end of the evaluation we put
$\varepsilon=1$), {we write} the solution to
(\ref{eq:dmrwa-i})--(\ref{eq:dmrwa-iii}) and the light-pressure force
(\ref{eq:frwa}) in the form
\begin{eqnarray}
    \varrho_{11}&=\sum\limits_{n=0}^{\infty}\varepsilon^{n}\varrho_{11}^{(n)},\label{eq:pert-i}\\
    \sigma_{12}&=\sum\limits_{n=0}^{\infty}\varepsilon^{n}\sigma_{12}^{(n)},\label{eq:pert-ii}\\
    F&=\sum\limits_{n=0}^{\infty}\varepsilon^{n}f^{(n)}.\label{eq:pert-iii}
\end{eqnarray}
From equations (\ref{eq:dmrwa-i})--(\ref{eq:dmrwa-iii}) it is easy to
see that
\begin{eqnarray}
  \label{eq:zero-i}
    \varrho_{11}^{(0)}=1, \quad \varrho_{11}^{(1)}=0, \\
  \label{eq:zero-ii}  \sigma_{12}^{(0)}=\sigma_{12}^{(2)}=0, \quad f^{(0)}=f^{(1)}=0.
\end{eqnarray}
The first order solution to (\ref{eq:dmrwa-i})--(\ref{eq:dmrwa-iii})
is
\begin{eqnarray}
  \label{eq:pert-i-s-i}
 \sigma_{12}^{(1)}(t)&=&\sigma_{12}^{(1)}(0)e^{-\frac{1}{2}\gamma t}+\frac{i}{2}\int\limits_{0}^{t}
    e^{\frac{1}{2}\gamma (t'-t)}\nonumber\\
&\times&    e^{-i\delta  t'} \left[ \Omega_{1}(t')e^{-i kz'}
+\Omega_{2}(t')e^{i kz'}  \right]d t',\\
  \label{eq:pert-i-s-ii}
    \sigma_{21}^{(1)}(t)&=&\sigma_{21}^{(1)}(0)e^{-\frac{1}{2}\gamma t}-\frac{i}{2}\int\limits_{0}^{t}
e^{\frac{1}{2}\gamma (t'-t)}\nonumber\\
&\times&e^{i\delta t'}\left[\Omega_{1}^{*}(t')e^{i kz'}
      +\Omega_{2}^{*}(t')e^{-i kz'}
    \right]d t',
\end{eqnarray}
where $z'$ is the coordinate of the atom at time $t'$. 
{The} second order term $f^{(2)}$ of the light-pressure force {is}
\begin{eqnarray}
  \label{eq:pert-f-ii}
   f^{(2)}&=&-\frac{i}{2}\hbar k \left[
      \Omega_{1}^{*}\sigma_{12}^{(1)}e^{i\delta t+i kz}-
      \Omega_{1}\sigma_{21}^{(1)}e^{-i\delta t-i kz}
    \right]\nonumber\\
   & +&\frac{i}{2}\hbar k \left[
      \Omega_{2}^{*}\sigma_{12}^{(1)}e^{i\delta t-i kz}-
      \Omega_{2}\sigma_{21}^{(1)}e^{-i\delta t+i kz} \right].
\end{eqnarray}
Substituting (\ref{eq:pert-i-s-i})--(\ref{eq:pert-i-s-ii}) in
(\ref{eq:pert-f-ii}) and averaging force over the wavelength, we
obtain the steady-state value of the light-pressure force
\begin{equation}
  \label{eq:pert-f-ii-res}
 \tilde{f}^{(2)}=
\frac{-2\hbar k\Omega_{0}^{2}\left(
     \frac{1}{2}\gamma+G
  \right)\delta kv
}{
  \left[\left(
     \frac{1}{2}\gamma+G
  \right)^{2}\!+\!(\delta +kv)^{2}    
  \right]\!
\left[\left(
     \frac{1}{2}\gamma+G
  \right)^{2}\!+\!(\delta -kv)^{2}    
  \right]
}.
\end{equation}
Here we assumed $z=z_{0}+vt$, $z'=z_{0}+vt'$, where $v$ is the
velocity of the atom, and took into account that, according
to~(\ref{eq:E-chaotic_re}) and (\ref{eq:E-chaotic_im}),
\begin{eqnarray}
\label{eq:Omega-chaotic-i}
  \langle\mathop{\mathrm{Re}} \Omega(t)\mathop{\mathrm{Re}} \Omega(t')\rangle
  &=&\frac{1}{2}\Omega_{0}^{2} e^{-G|t-t'|},\\
  \langle\mathop{\mathrm{Im}} \Omega(t)\mathop{\mathrm{Im}}\Omega(t')\rangle 
  &=&\frac{1}{2}\Omega_{0}^{2}
      e^{-G|t-t'|},
\label{eq:Omega-chaotic-ii}
\end{eqnarray}                          
where $\Omega_{0}=\sqrt{\langle|\Omega|^{2}\rangle}$. 

Equation (\ref{eq:pert-f-ii-res}) shows that the damping force occurs for
$\delta>0$, when the carrier frequency $\omega$ of the the
counter-propagating waves is less then the transition frequency
$\omega_{0}$ in the atom. This corresponds with a well-known result
for Doppler cooling of atoms by a monochromatic standing
wave~\cite{Metcalf}.

Similar calculations {give} the second-order solution to
(\ref{eq:dmrwa-i})--(\ref{eq:dmrwa-iii}). {When} averaged over the
wavelength, in the case of slow atoms ($k|v|\ll\gamma$), {this} gives
\begin{equation}
  \label{eq:pert-ii-av-res-d}
    \tilde{\varrho}_{22}^{(2)}=\frac{\Omega_{0}^{2}\left(
        \frac{1}{2}\gamma+G
      \right)}{\gamma
      \left[
          \left(
        \frac{1}{2}\gamma+G
      \right)^{2}+\delta^{2}    \right]}.
\end{equation}
Knowledge of the population of the excited state allows {one} to find
the rate of spontaneous emission of photons
$\gamma \tilde{\varrho}_{22}$, and, {from this to} calculate the
momentum diffusion coefficient~\cite{Ada97}
\begin{equation}
  \label{eq:Dp-27}
D_{p}=\frac{1}{2}(\hbar k)^{2}(1+Q+\xi)\gamma \tilde{\varrho}_{22}.
\end{equation}
{This coefficient quantifies} the time dependence of the mean-squared momentum
deviation $p$ from the mean value $\langle p\rangle$,
\begin{equation}
  \label{eq:dp}
  \langle(p-\langle p\rangle)^{2}\rangle=2D_{p}t.
\end{equation}
{In~(\ref{eq:Dp-27})} the Mandel parameter $Q$ takes into account
non-Poisson statistics of scattered photons. For low intensities
this term is small and may be neglected. Parameter $\xi$ is
determined by the angular distribution of the scattered photons. For
an one-dimensional model considered here $\xi=1$ (photons are
emitted in two opposite directions along the propagation of the light
rays). It is for this model the well-known formula for the minimal
temperature of the two-level atoms in the field of the standing wave,
\begin{equation}
  \label{eq:TD}
T_{D}=\frac{\hbar \gamma}{2k_{B}},  
\end{equation}
where $k_{B}$ is the Boltzmann constant, was derived.

Taking into account that the friction force, which slows the atom's
velocity down, according to~(\ref{eq:pert-f-ii-res}) is equal to
\begin{equation}
  \label{eq:friction}
\tilde{f}^{(2)}=-\alpha v  
\end{equation}
for $v\ll\gamma/k$, where
\begin{equation}
  \label{eq:alpha}
\alpha =
\frac{2\hbar k^{2}\Omega_{0}^{2}\left(
     \frac{1}{2}\gamma+G
  \right)\delta 
}{
  \left[\left(
     \frac{1}{2}\gamma+G
  \right)^{2}\!+\!\delta^{2}    
  \right]^{2}
},
\end{equation}
we calculate the temperature of the atomic ensemble~\cite{Ada97}
\begin{equation}
T =\frac{D_{p}}{\alpha k_{B}}=\frac{\hbar}{2k_{B}}
\left(
  G+\frac{\gamma}{2}
\right)
\left[
  \frac{G+\frac{\gamma}{2}}{\delta}+\frac{\delta}{G+\frac{\gamma}{2}}
\right].
  \label{eq:T}
\end{equation}
The temperature (\ref{eq:T}) reaches the minimum value
\begin{equation}
  \label{eq:Tmin}
T_{\min} =\frac{D_{p}}{\alpha k_{B}}=\frac{\hbar}{k_{B}}
\left(
  G+\frac{\gamma}{2}
\right)
\end{equation}
when the detuning equals
\begin{equation}
  \label{eq:opt}
\delta=\delta_{\mathrm{opt}} =\left(
  G+\frac{\gamma}{2}
\right).
\end{equation}
Formula (\ref{eq:Tmin}), when $G\to0$, coincides with formula
(\ref{eq:TD}) for the minimal temperature of the atoms in the field of
a standing monochromatic wave.

\section{Numerical calculation procedure}
\label{sec:numer-calc-proc}


The time evolution of the statistical properties of the atomic
ensemble is calculated by the Monte Carlo wave-function
method~\cite{Mol93} with the subsequent ensemble averaging of the
first and the second moments of the velocity and coordinate
distribution functions. The state vector of the atom
\begin{equation}
  \left| \psi\right\rangle=
  c_{1}\left| 1\right\rangle+c_{2}\exp({-i\omega_{0}t})\left| 2\right\rangle,
\label{eq:psi}
\end{equation}
is determined from the Schr\"odinger equation
\begin{equation}
i\hbar\frac{d}{dt}\left| \psi\right\rangle=H\left| \psi\right\rangle
\label{eq:Schrod}
\end{equation}
with the simulation of a quantum jump probability after every step of
the integration.

The Hamiltonian in the Schr\"odinger equation~(\ref{eq:Schrod}) is
\begin{equation}
{{H}}=H_{0}+H_{\mathrm{int}}+H_{\mathrm{rel}},
\label{eq:Ham}
\end{equation}
where
\begin{equation}
H_{0}= \hbar\omega_{0}|2\rangle\langle2|
\end{equation}
describes the atom in the absence of fields and relaxation. {The term}
\begin{equation}
  H_{\mathrm{int}}= -{\mathbf{d}}_{12}|1\rangle\langle2|{\mathbf{E}}(t)-
  {\mathbf{d}}_{21}|2\rangle\langle1|{\mathbf{E}}(t),
\label{eq:Hint}
\end{equation}
is responsible for the interaction of the atom with the field, and
\begin{equation}
H_{\mathrm{rel}}= -\frac{i\hbar\gamma}{2}|2\rangle\langle{}2|                         
\label{eq:Hrel}
\end{equation}
describes the relaxation due to the spontaneous emission.  Because of
this term the Hamiltonian (\ref{eq:Ham}), describing a closed
two-level system, is non-Hermitian, contrary to the Hamiltonian in the
equation for the density matrix.

Substitution~(\ref{eq:psi}) and (\ref{eq:Ham}) in (\ref{eq:Schrod})
gives the equation for $c_{1}$ and $c_{2}$. After applying the
rotating wave approximation (RWA)~\cite{Shore} we finally arrive at
\begin{eqnarray}
\frac{d}{d t}c_{1}&=&-\frac{i}{2}\left[\Omega_{1}e^{-i kz}+
    \Omega_{2}e^{i kz}\right]c_{2}e^{-i\delta t},\label{eq:dci}\\
\frac{d}{d t}c_{2}&=&-\frac{i}{2}\left[\Omega_{1}^{*}e^{i kz}+
    \Omega_{2}^{*}e^{-i kz}\right]c_{1}e^{i\delta t}-\frac{\gamma}{2}c_{2}.
\label{eq:dcii}
\end{eqnarray} 
We simulate the stochastic fields $\Omega_{1}, \Omega_{2}$ by {an}
Ornstein-Uhlenbeck (OU) process $\Xi(t)$~\cite{Horsthemke} with the
autocorrelation function
\begin{equation}
\langle\Xi(t)\Xi(t')\rangle=BGe^{-G|t-t'|}.
\label{eq:OU}
\end{equation} 
This process is the solution to the Langevin differential equation
\begin{equation}
  \label{eq:Langevin}
  \frac{d}{dt}\Xi(t)=-G\Xi(t)+G\xi(t).
\end{equation}
Here $\xi(t)$ is the Gaussian white noise with the
autocorrelation function
\begin{equation}
  \label{eq:white}
\langle\xi(t_{1})\xi(t_{2})\rangle  =2B\delta(t_{1}-t_{2})
\end{equation}
and $\delta(t)$ is the Dirac delta function.

The stochastic fields $\Omega_{1}(t), \Omega_{2}(t)$ {are} related to
$\Xi(t)$ by the expressions
\begin{equation}
\label{eq:Omega-Xi}
\Omega_{1}(t)=\Xi(t-z/c),\quad\Omega_{2}(t)=\Xi(t+z/c).
\end{equation}

{The right-hand side} of~(\ref{eq:OU}) coincides with {the right-hand
  sides} of (\ref{eq:Omega-chaotic-i}), (\ref{eq:Omega-chaotic-ii})
for $\Omega_{0}=\sqrt{2BG}$.  Function $\Xi(t)$ is generated by the
algorithm of Fox and others~\cite{Fox,Vemuri}, as was done
in~\cite{Romanenko-OC,Kuhn,Yatsenko}.  The value of $\Xi(t_{j+1})$ at
discrete times $t_{j}=t_{j-1}+\Delta t$ is calculated from
$\Xi(t_{j})$ according to
\begin{equation}\label{eq:noise}
\Xi (t_{j+1})=\Xi(t_j)e^{-G\Delta{}t}+h(t_j).
\end{equation}
Here $h(t_j)$ obeys Gaussian statistics with a zero first moment and
the second moment given by
\begin{equation}\label{eq:gauss}
\langle h(t_j)^2 \rangle  =   B G
\left(1-e^{-2G\Delta{}t}\right).
\end{equation}

{The} state vector $|\psi(t)\rangle$ is normalized to unity at time
$t$. {Because} the Hamiltonian~(\ref{eq:Ham}) is not Hermitian, after
time step $\Delta t$ the state vector
$|\psi^{(1)}(t+\Delta{}t)\rangle$ is not normalized to
unity. According to~\cite{Mol93},
\begin{equation}
  \Delta{}P=1-\langle\psi^{(1)}(t+\Delta{}t)|\psi^{(1)}
  (t+\Delta{}t)\rangle
\label{eq:phiIN}
\end{equation} 
is the probability of a quantum jump. If quantum jump occurs,
the state vector becomes
\begin{equation}
|\psi(t+\Delta{}t)\rangle=|1\rangle.
\end{equation}
In the opposite case we normalize the state vector and it becomes
\begin{equation}
  |\psi(t+\Delta{}t)\rangle=\frac{|\psi^{(1)}(t+\Delta{}t)
    \rangle}{\sqrt{1-\Delta{}P}}.
\end{equation}
To simulate a quantum jump, we introduce the random variable
$\epsilon$, which is uniformly distributed between zero and one. If it
is smaller than $\Delta{}P$, the quantum jump occurs.

The elements of the density matrix in~(\ref{eq:F}) are expressed in
terms of the probability amplitudes $c_{1}$ and $c_{2}$ by
\begin{equation}
  \varrho_{12} =c_{1}c_{2}^{*} e^{i\omega_{0}t},\quad
  \varrho_{21} =c_{2}c_{1}^{*} e^{-i\omega_{0}t}.
\label{eq:cc}
\end{equation}

The light-pressure force~(\ref{eq:F}) {averaged} over the period of
the field's oscillations $2\pi/\omega_{0}$ reads
\begin{equation}
  F= \hbar{}k\mathop{\mathrm{Im}}\left\{c_{1}c_{2}^{*}e^{i\delta t}
    \left[\Omega_{1}^{*}e^{ikz}-\Omega_{2}^{*}e^{-ikz}\right]
  \right\}\left(
 |   c_{1}|^{2}+|c_{2}|^{2}
  \right)^{-1} .
  \label{eq:FF}
\end{equation}

To simulate the atom's motion, we simultaneously solve
{equations}~(\ref{eq:ma}) and (\ref{eq:dci}), (\ref{eq:dcii}), where
the light-pressure force is given by~(\ref{eq:FF}).  In addition, the
atomic momentum is stochastically changed due to the spontaneous
emission and fluctuations of the absorption and the stimulated
radiation emission process~\cite{Minogin}.  In our calculations, for
the sake of simplicity, we accept that the atomic momentum changes by
$\pm\hbar k$ with an identical probability in the course of
spontaneous emission.

In the case of {weak} laser radiation intensity {and negligible}
population of the excited state, the lightpressure force and the
momentum diffusion coefficient are equal to the sums of corresponding
quantities for each of the counter-propagating waves~\cite{Mol91}. The
spontaneous emission occurs after each absorption of a photon, and the
fluctuations of the momentum change in the stimulated processes occur
as often as the similar fluctuations in spontaneous
radiation~\cite{Minogin} (see Appendix). {We applied this model} to
the computer simulation of the fluctuation-driven momentum variation
in the process of the interaction of atoms with the field of the
counter-propagating low-intensity pulses~\cite{Romanenko-PRA} and with
the field of the counter-propagating bichromatic
waves~\cite{Romanenko-UJP-2016}.

In the case of counter-propagating waves of high intensity, which is
required to reduce the size of the atomic cloud in a trap, we may
consider the momentum diffusion in the stimulated processes in a
similar way, bearing in mind that the result should be treated as an
estimation. A probable error can be associated with larger
fluctuations of the momentum change owing to stimulated radiation
processes in comparison with those occurring owing to the spontaneous
radiation emission. {The resulting calculation underestimates} the
atomic cloud size and the atomic temperature in the trap.

In summary, the calculation algorithm of the motion of atoms in the
field is as follows. {Equations}~(\ref{eq:ma})
and~(\ref{eq:dci}),~(\ref{eq:dci}) are integrated by the
{fourth-order} Runge-Kutta method of the fourth order.  After every
step we randomly decide whether a quantum jump occurs, and the state
vector is {renormalized}.  If the quantum jump occurs, the velocity of
the atom changes by
\begin{equation}
  \Delta{}v=\hbar{}k(\epsilon_1-0.5)/(M|\epsilon_1-0.5|)+\hbar{}k(\epsilon_2-0.5)/(M|\epsilon_2-0.5|),
\end{equation}
where $\epsilon_{1,2}$ are random numbers uniformly distributed over
the interval [0,1]. One of the terms simulates the momentum
fluctuation due to a spontaneously emitted photon, and the other
simulates the momentum fluctuation resulting from fluctuations of the
stimulated absorption and emission of photons.  The Ornstein-Uhlenbeck
process, which describes the time dependence of the radiation field of
the laser, is simulated by the equations~(\ref{eq:noise}) and
(\ref{eq:gauss}).

After simulation of the motion of each atom in the ensemble we find
the time dependence of the average velocity and the average coordinate
of the atoms and the square root of the mean square deviation of the
velocity and the coordinate from the corresponding average values.

\section{Results and discussion}
\label{sec:results-discussion}

It is well known that the cycling atom-field interaction can be
realized between two states of some atoms~\cite{Metcalf}. Our
calculations was carried out for $^{23}\mathrm{Na}$ atoms, in which
the cyclic interaction can be realized for the transition
$3{} ^{2}S_{1/2} - 3{} ^{2}P_{3/2}$. The transition wavelength is
$\lambda=589.16$~nm, the rate of spontaneous emission is
$\gamma=2\pi\times10$~MHz, the Doppler cooling temperature limit for
sodium atoms is $T_{D}=240$~$\mu$K~\cite{Metcalf}.

The motion of the atom in a stochastic field {depends substantially}
on the field parameters. Among the whole set of parameters, we are
interested in those that provide the motion of atoms in a narrow
interval of coordinates $z$ at the vicinity of the trap center.

\begin{figure} 
\centerline{\includegraphics[]{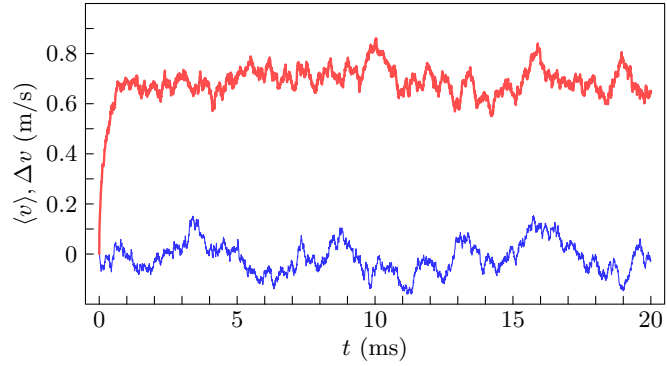}}
\caption{The time dependence of the average velocity
  $\langle v\rangle$ (thin line) and the standard deviation of the
  velocity from its average value $\Delta v$ (thick line) of the
  atomic ensemble of 100 sodium atoms for
  $\Omega_{0}=2\pi\times2$~MHz, $\delta=2\pi\times20$~MHz,
  $G=2\pi\times20$~MHz, $B=2\pi\times0.1$~MHz, initial velocity
  $v_{0}=0$}
\label{fig-4}
\end{figure}

To test the method of calculation of the statistical parameters of the
atomic ensemble we simulated the motion of sodium atoms in a weak
field and compared the results with the analytical formula for the
temperature of the atomic ensemble~(\ref{eq:T}).
\begin{figure} 
\centerline{\includegraphics[]{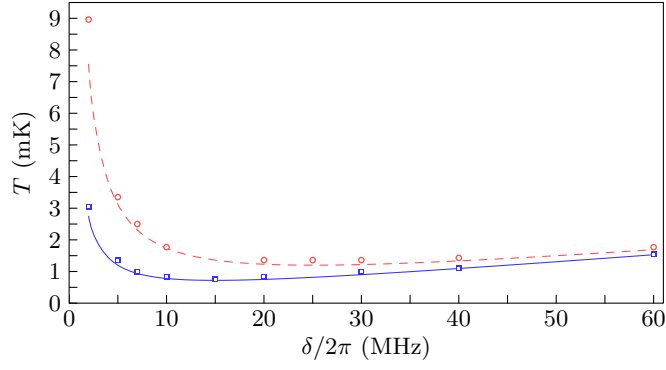}}
\caption{The dependence of the temperature of the atomic ensemble of
  100 sodium atoms on the detuning $\delta$ of the carrier frequency
  of the stochastic waves $\omega$ from the atomic transition
  frequency $\omega_{0}$ for $\Omega_{0}=2\pi\times2$~MHz after 20 ms
  of the interaction of the atoms with the field. Squares and circles
  are the results of numerical calculation, lines show the calculation
  according to formula~(\ref{eq:T}). Squares and solid curve correspond
  to $G=2\pi\times10$~MHz, $D=2\pi\times0.2$~MHz, circles and dashed
  line correspond to $G=2\pi\times20$~MHz, $B=2\pi\times0.1$~MHz}
\label{fig-3}
\end{figure}
Figure \ref{fig-3} shows a good agreement of numerical calculations of
the temperature of the atomic ensemble with expression~(\ref{eq:T}).
Figure \ref{fig-4} shows an example of the time dependence of the
average velocity and the standard deviation of the velocity from its
mean value for the parameters corresponding to one of the points in
Figure \ref{fig-3}.

\begin{figure} 
\centerline{\includegraphics[]{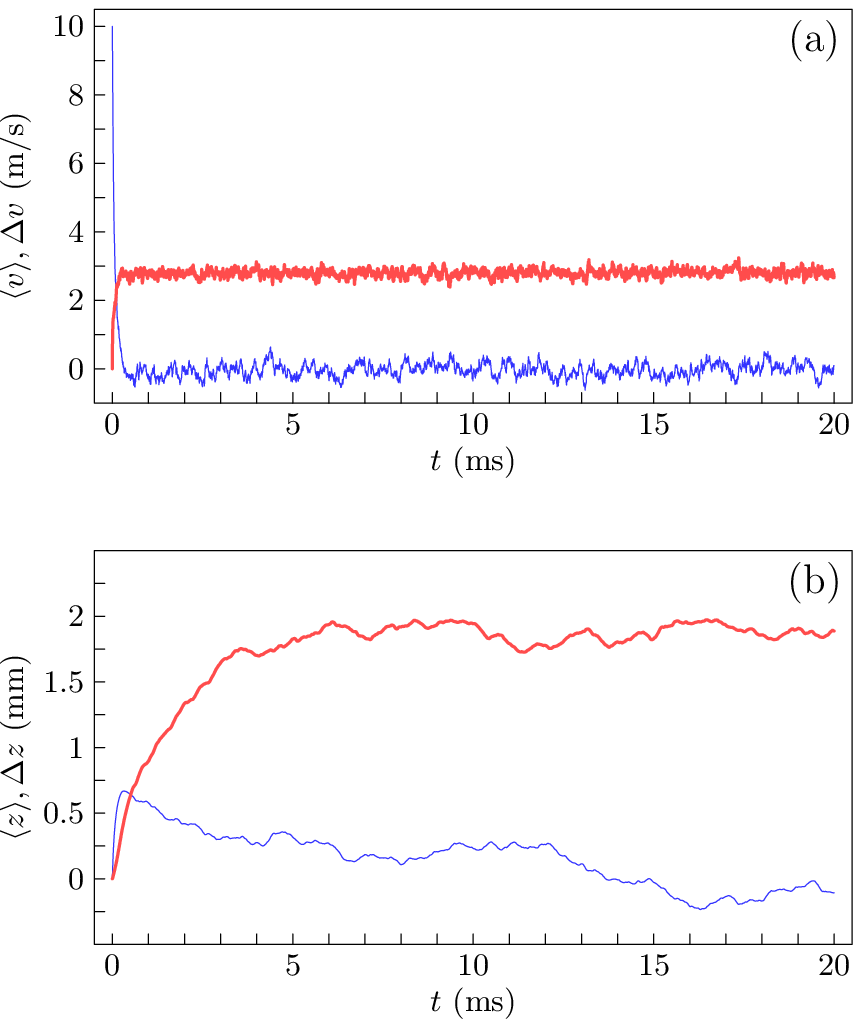}}
\caption{(a) The time dependence of the average velocity
  $\langle v\rangle$ (thin line) and the standard deviation of the
  velocity from the average value $\Delta v$ (thick line) for the
  atomic ensemble. (b) The time dependence of the average coordinate
  $\langle z\rangle$ (thin line) and the standard deviation of the
  coordinate from the average value $\Delta z$ (thick line) for the
  same atomic ensemble. Parameters: $\Omega_{0}=2\pi\times50$~MHz,
  $\delta=2\pi\times15$~MHz, $G=2\pi\times20$~MHz,
  $B=2\pi\times62.5$~MHz. Calculation was carried out for 200 sodium
  atoms, initial velocity $v_{0}=10$~m/s}
\label{fig-5}
\end{figure}

Over time, the size of the atomic cloud increases until the light
pressure force prevents atoms from moving away from the center of the
trap. Because the light-pressure force is small when the intensity of
the laser beam is low, the atomic cloud does not stop expanding even
for $t>100$~ms. Confinement of the atoms in the trap can be achieved
by increasing the intensity of the laser radiation, as is shown in
Figure~\ref{fig-5}.
Atoms with the initial velocity $v_{0}=10$~m/s slow down during
0.35~ms, and then their velocity fluctuates near zero with the
variance of $\Delta v=2.8$~m/s, which corresponds to temperature
$T=0.021$~K (see Figure \ref{fig-5}(a)). The quasi-stationary
distribution of atoms in space with mean square deviation from the the
center of the trap $\Delta z=2$~mm is reached after 6~ms. The center
of mass of the ensemble slows down and then tends to the origin of the
coordinates (the center of the trap), where it fluctuates (see
Figure~\ref{fig-5}(b)).

\begin{figure}
\centerline{\includegraphics[]{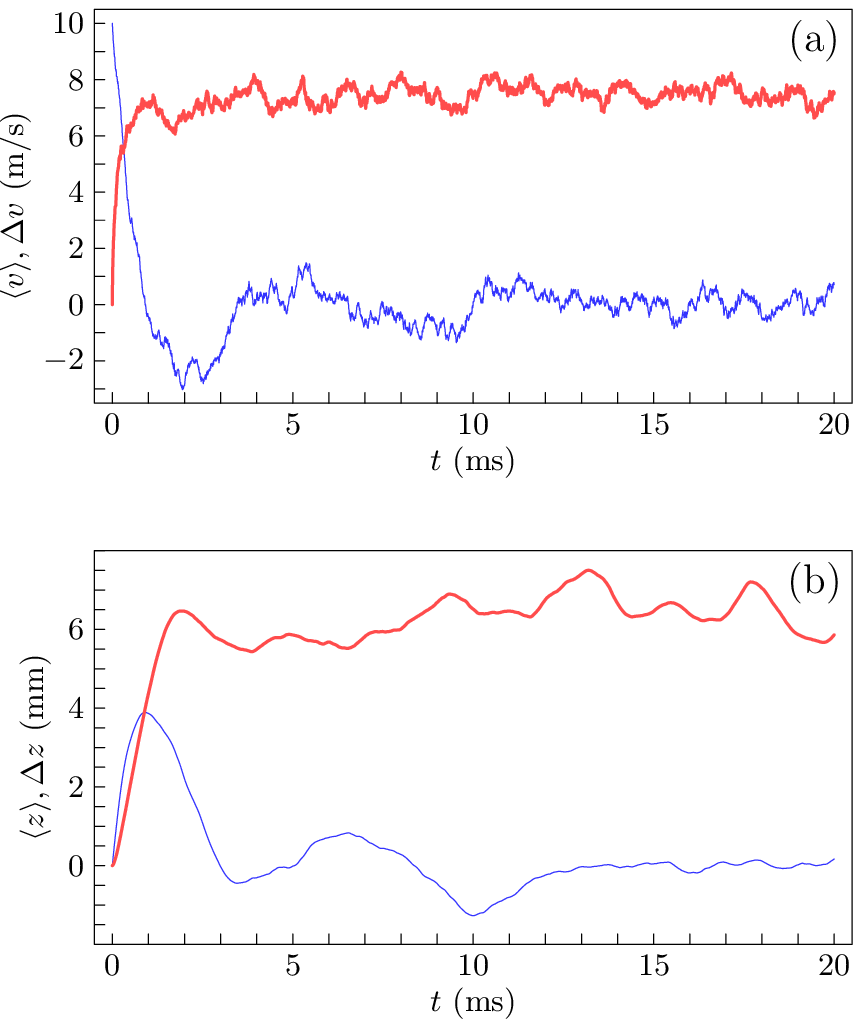}}
\caption{(a) The time dependence of the average velocity
  $\langle v\rangle$ (thin line) and the standard deviation of the
  velocity from the average value $\Delta v$ (thick line) for an
  atomic ensemble. (b) The time dependence of the average coordinate
  $\langle z\rangle$ (thin line) and the standard deviation of the
  coordinate from the average value $\Delta z$ (thick line) for the
  same atomic ensemble. Parameters: $\Omega_{0}=2\pi\times50$~MHz,
  $\delta=2\pi\times1$~MHz, $G=2\pi\times20$~MHz,
  $B=2\pi\times62.5$~MHz. Calculation was carried out for 200 sodium
  atoms, initial velocity $v_{0}=10$~m/s}
\label{fig-6}
\end{figure}

When the detuning $\delta$ becomes small, the friction force becomes
less, and the decaying movement to the center of the trap changes to
oscillation with the amplitude gradually decreasing to zero. An
example of such a movement is shown in Figure \ref{fig-6}. The root
mean square value of the velocity at low detunings grows, resulting in
rising temperature of the atomic ensemble, which is 0.15~K for Figure
\ref{fig-6}.
Reduced friction force also explains approximately five times more
deviation of $\langle z\rangle$ from the center of the trap at the
beginning of the movement of atoms in Figure~\ref{fig-6}(b) compared to
Figure \ref{fig-5}(b).

\begin{figure}
\centerline{\includegraphics[]{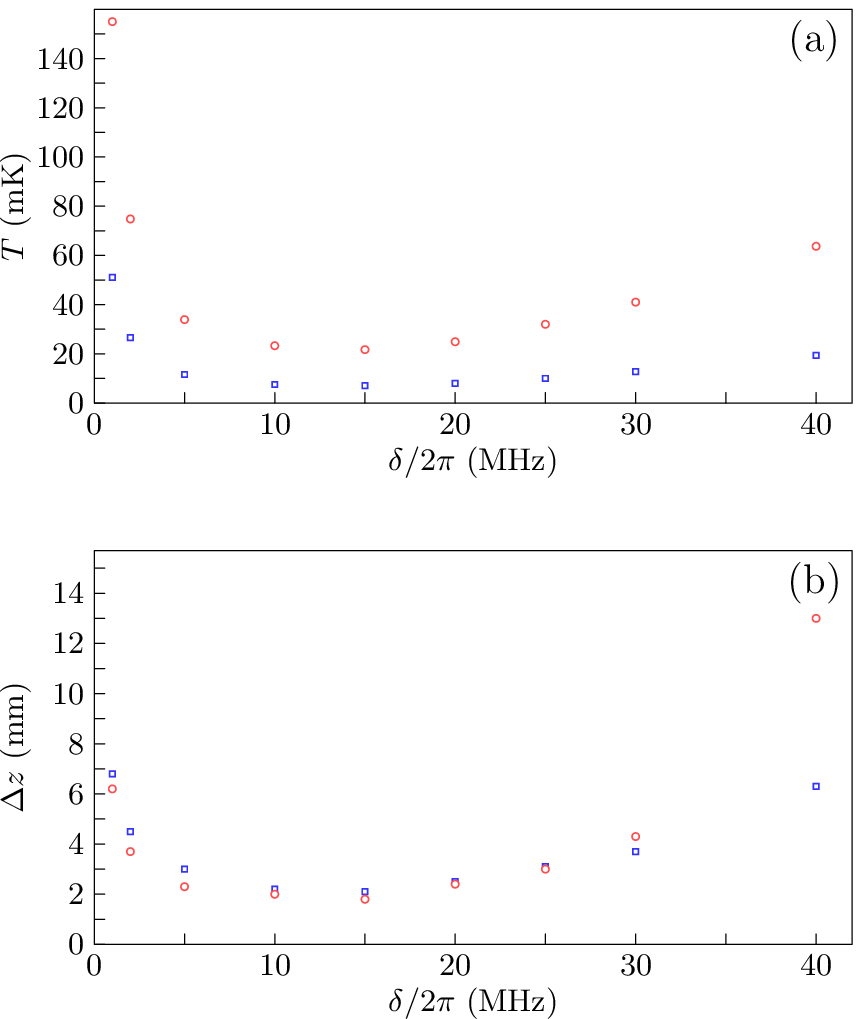}}
\caption{The dependence of (a) the temperature $T$ of the atomic
  ensemble and (b) the standard deviation $\Delta z$ of sodium atoms
  from the center of the trap on the detuning $\delta$. Parameters:
  $G=2\pi\times20$~MHz, squares --- $\Omega_{0}=2\pi\times20$~MHz,
  $B=2\pi\times10$~MHz, circles --- $\Omega_{0}=2\pi\times50$~MHz,
  $B=2\pi\times62.5$~MHz. Calculation was carried out for 200 sodium
  atoms }
\label{fig-7}
\end{figure}

Figure \ref{fig-7} shows the dependence of the temperature $T$ and the
standard deviation $\Delta z$ of the sodium atoms from the center of
the trap on the detuning $\delta$ for high intensity of the laser
radiation ($\Omega_{0}\gg\gamma$).
As can be seen, the minimal temperature and the size of the atomic
cloud is achieved when $\delta=2\pi\times15$~MHz, which is
significantly below $2\pi\times25$~MHz given by the
formula~(\ref{eq:opt}). The minimal {temperatures} of the atomic
ensemble for $\Omega_{0}=2\pi\times20$~MHz and
$\Omega_{0}=2\pi\times50$~MHz are $7.1$~mK and $21.7$~mK. {These
  significantly exceed by} $1.2$~mK the value given by
formula~(\ref{eq:Tmin}).  The temperature of sodium atoms and the size
of the atomic cloud ($2\Delta z$) in the trap considered here {are}
significantly greater (approximately 30 and 12 times) {than}
corresponding values for Doppler cooling and trapping of sodium atoms
in the widely used magneto-optical trap~\cite{Raab1987}.
Nevertheless, {our results show} that even highly disordered field of
the counter-propagating stochastic waves can form a trap for atoms. We
believe that the physical nature of this trap is a correlation of the
counter-propagating wave with {the} time dependent envelope, which is
significant for distances less then $c/G$ from the center of the trap.
Taking into account the possible formation of the trap by the trains
of counter-propagating pulses~\cite{Romanenko-PRA} and the bichromatic
counter-propagating waves~\cite{Romanenko-UJP-2016}, we suggest a
hypothesis that any polychromatic counter-propagating waves with
discrete spectrum or waves described by a stationary stochastic
process, one of which repeats the other, can form a trap for atoms.

\section{Conclusions}
\label{sec:conclusions}

The simulation of the motion of atoms in the field of
counter-propagating stochastic waves (the chaotic-field model), one of
which repeats the other, shows that the atoms can be confined in a
small area (a few mm) near the point where the amplitudes and phases
of the counter-propagating waves are the same. The minimal temperature
of the atoms in low-intensity counter-propagating waves is determined
by the rate of spontaneous emission of photons by the atoms in the
excited state and the autocorrelation time of the stochastic
process. If the intensity of the waves is large, it significantly
affects the temperature of the atoms.

We believe that the trap formed the stochastic waves {and the} trap
formed by the counter-propagating pulses or bichromatic waves {have a}
common physical basis, {the} correlation of the counter-propagating
waves. This point of view leads to the suggestion {that} any
polychromatic counter-propagating waves with discrete spectrum or
waves described by a stationary stochastic process, one of which
repeats the other, can form a trap for atoms.

{Because} the temperature of the atomic ensemble in low-intensity
counter-propagating stochastic waves is determined by the width of the
spectrum of the stochastic field and the rate of spontaneous emission,
such fields can be used for the formation of ensembles of atoms with a
given (higher than Doppler cooling limit) temperature.

\section{Acknowledgements}

Funding: This work was supported by the State Fund for Fundamental
Research of Ukraine [grant number F64/26-2016]; National Academy of
Sciences of Ukraine in the frame of the goal-oriented complex program
of fundamental researches ``Fundamental issues in creation of new
nanomaterials and nanotechnologies'' [grant number 3/16-H).
We thank B. W. Shore for careful reading of the manuscript.


\appendix

\section{Momentum diffusion of atoms in the field of one travelling
  wave}

Here we explain the origin of the momentum diffusion of atoms in the
field of laser radiation for an example of one travelling wave.  We
follow the book~\cite{Minogin}, adapting the presentation for
one-dimensional model considering here.

We assume that the atom's momentum at the time instant $ t $ is
$ \mathbf{p}_{0} $. After time $ \Delta t $ it becomes
\begin{equation}
\mathbf{p}=\mathbf{p}_{0}+\hbar{}\mathbf{k}(N_{+}-N_{-})-\sum_{s}\hbar{}\mathbf{k}_{s}.
\label{eq:p}
\end{equation}
The second term describes the change of momentum due absorption and
stimulated emission. The wave vector $ \mathbf{k} $ is directed along
$ z $-axis, $N_{+}$ and $N_{-}$ are the numbers of the absorbed and
emitted photons.  The third term in~(\ref{eq:p}) gives the momentum
change due to the spontaneous emission of the photons with the wave
vectors $ \mathbf{k}_{s} $. For the one-dimensional model we assume
$\mathbf{k}_{s}=\pm\mathbf{k}$ with equal probability and summation is
over all the spontaneously emitted photons during time $\Delta t$.

After averaging~(\ref{eq:p}) over an ensemble of atoms we find
\begin{equation}
\langle\mathbf{p}\rangle=\langle\mathbf{p}_{0}\rangle+\hbar{}\mathbf{k}(\langle N_{+}\rangle-\langle N_{-}\rangle),
\label{eq:pav}
\end{equation}
where $ \langle\mathbf{p}_{0}\rangle $ is the initial average
momentum, $ \langle N_{+}\rangle $ is the average number of the
absorbed photons, $ \langle N_{-}\rangle $ is the average number of
the photon emitted due to stimulated emission. Here we took into account that
\begin{equation}
\Bigl\langle\sum_{s}^{}\mathbf{k}_{s}\Bigr\rangle=0.
\label{eq:ksav}
\end{equation}
Subtracting (\ref{eq:p}) from (\ref{eq:pav}) we find the momentum
fluctuation,
\begin{equation}
\Delta \mathbf{p} =\mathbf{p}-\langle\mathbf{p}\rangle=
(\mathbf{p}_{}-\langle\mathbf{p}_{0}\rangle)+\hbar\mathbf{k}\Delta N_{i}-\sum_{s}^{}\hbar\mathbf{k}_{s},
\label{eq:Dp} 
\end{equation}
where $\Delta N_{i} = N_{i}-\langle N_{i} \rangle$ is the variation of
the difference $ N_{i}=N_{+}-N_{-} $ from the ensemble average value.

The average square of the momentum fluctuations along $ z $-axis is 
\begin{equation}
\langle\Delta {p}_{z}^{2}\rangle =\langle\Delta {p}_{0z}^{2}\rangle+
\hbar^{2}{k}^{2}\langle(\Delta N_{i})^{2}\rangle+\hbar^{2}{k}^{2}\langle N_{s}\rangle.
\label{eq:DpDp} 
\end{equation}
Here $ \langle N_{s}\rangle $ is the average number of the
spontaneously emitted photons $\langle\Delta {p}_{0z}^{2}\rangle$ is
the initial momentum spreading,
$\hbar^{2}{k}^{2}\langle(\Delta N_{i})^{2}\rangle$ arises because of
the stimulated processes (absorption and emission),
$\hbar^{2}{k}^{2}\langle N_{s}\rangle$ is due to spontaneous emission.

Assuming the Poisson photons statistics, we find
\begin{equation}
\langle(\Delta N_{i})^{2}\rangle=\langle N_{i}\rangle.
\label{eq:Poisson}
\end{equation}
Noting that $ \langle N_{i}\rangle=\langle N_{s}\rangle $, we finally get
\begin{equation}
\langle\Delta {p}_{z}^{2}\rangle =\langle\Delta {p}_{0z}^{2}\rangle+
\hbar^{2}{k}^{2}\langle N_{s}\rangle+\hbar^{2}{k}^{2}\langle N_{s}\rangle.
\label{eq:DpDpf} 
\end{equation}
This equation can be interpreted physically as changing of atomic
momentum by $\pm\hbar k$ due to spontaneous emission and by
$\pm\hbar k$ due to stimulated process (emission an absorption) every
spontaneous emission event.


\sloppy


\end{document}